\documentclass[a4paper]{article}

\usepackage{INTERSPEECH2021}
\usepackage{multirow}
\usepackage{subcaption}

\title{On-Device Personalization of Automatic Speech Recognition Models for Disordered Speech}
\name{Katrin Tomanek, Fran{\c{c}}oise Beaufays, Julie Cattiau, Angad Chandorkar, Khe Chai Sim}
\address{Google, USA}
\email{\{katrintomanek,fsb,juliecattiau,angadc,khechai\}@google.com}

\begin{document}

\maketitle
\begin{abstract}
While current state-of-the-art Automatic Speech Recognition (ASR) systems achieve high accuracy on typical speech, they suffer from significant performance degradation on disordered speech and other atypical speech patterns. Personalization of ASR models, a commonly applied solution to this problem, is usually performed in a server-based training environment posing problems around data privacy, delayed model-update times, and communication cost for copying data and models between mobile device and server infrastructure. In this paper, we present an approach to on-device based ASR personalization with very small amounts of speaker-specific data. We test our approach on a diverse set of 100 speakers with disordered speech and find median relative word error rate improvement of 71\% with only 50 short utterances required per speaker. When tested on a voice-controlled home automation platform, on-device personalized models show a median task success rate of 81\%, compared to only 40\% of the unadapted models.
\end{abstract}

\section{Introduction} \label{sec:intro}

Accuracy of Automatic Speech Recognition (ASR) systems has improved significantly over recent years given more powerful model architectures and their training on tens of thousands of hours of speech data. While exhibiting state-of-the-art performance on a large variety of speakers, these models still perform poorly on speech and voices with very different characteristics. For speakers with a speech disorder -- such as dysarthria, deaf speech, severe hypernasality due to cleft lip and palate, or stuttering -- recognition accuracy is still unacceptably low~\cite{moore2018}, rendering the technology unusable for many of the speakers who could benefit the most. The poor recognition is due to the diversity and complexity of atypical speech patterns~\cite{darley1975motor}, and the lack of sufficient training data~\cite{caves2009}.
To navigate these challenges, a common solution is to fine-tune models originally optimized for typical speech on user-specific data. ASR model personalization has shown to work well for a variety of atypical speech patterns, including accents and speech disorders~\cite{Zhu2019,Shor2019,Gale2019,mustafa2014}.

As modern smart phones are becoming powerful enough to run offline ASR entirely on-device, this eliminates the need to send the user audio data to a server for inference. For example, the all-neural on-device ASR model based on the recurrent neural network transducer (RNN-T) architecture~\cite{schalkwyk2019,graves2012sequence,graves2013sequence} can be used to effectively combine the acoustic and language model components into a single compact neural network model and achieve high-quality large vocabulary continuous speech recognition performance~\cite{he2019streaming,young1996review}. 

From a privacy standpoint, it is desirable for such offline ASR models to also perform personalization on user devices without having to ever upload sensitive user voice data to a server for processing. In a scenario with frequent model adaptation, on-device training is a solution to reduce communication cost with a server, and it can offer a faster turnaround for users to experience improved personalized models. However, there are many challenges to perform on-device ASR personalization reliably: Memory and compute resources on mobile devices are limited, and the availability and reliability of on-device training data are also very different compared to server-side training, where data can be carefully collected and annotated. Some of these challenges are discussed in recent work~\cite{sim19interspeech,sim19asru}.

In this paper we present an approach to on-device personalization of end-to-end ASR models. In extensive experiments on speakers with impaired speech
we show that with as little as 50 recorded utterances, our proposed approach reduces the median word error rate (WER) by 71\% compared to readily available ASR models optimized for typical speech. We also show direct impact of these personalized ASR models on a voice-controlled application, i.e. home automation: on-device personalized ASR obtained a median task success rate of 81\%, compared to the substandard 40\% of the baseline model. Finally, we report insights from a small user study where an individual with severely disordered speech used our on-device personalization strategy on their mobile device.

\section{ASR Personalization}
\label{sec:ASR_pers_disordered_speech}

We use a modification of the RNN-T model architecture which allows deployment on mobile devices and supports streaming \cite{he2019streaming}. We use 8 LSTM encoder layers and 2 LSTM layers for the language model component.  Input features are 128-dimensional log Mel features computed every 10 milliseconds. 4 consecutive features are stacked with a stride of 3 frames to yield a 512-dimensional input to the encoder every 30 milliseconds. There are 4096 output units that correspond to the word piece tokens. We used the implementation described in~\cite{bagby2018efficient} to compute the RNN-T loss and gradients.

ASR personalization approaches can be categorized into language model (LM) and acoustic model (AM) adaptation. The former focusses on methods to learn user-specific LMs such as LM fusion~\cite{toshniwal2018comparison,gulcehre2015using,sriram2017cold} and personalized vocabulary such as biasing~\cite{aleksic2015improved,hall2015composition}. AM adaptation techniques~\cite{sim2017_springer,samarakoon2016factorized,xue2014singular,zhao2016low,saon2013speaker,senior2014improving,swietojanski2014learning,tan2015cluster,wu2015multi} attempt to reduce mismatch in terms of users' voice characteristics, such as accent, pitch and speaking rate. In this paper, our focus is on acoustic adaptation to improve ASR models for speakers with speech disorder.

\subsection{Server-side Personalization}
\label{ss:on_server_pers}

ASR personalization can be done via model-fine tuning, starting from a speaker-independent base model and adapting (all or parts of the) model weights to the specific target speaker's data. Fine-tuning is typically performed in a server environment, using accelerators, with relatively large amounts of memory being available, and with access to all adaptation data of a speaker at once. Training data is partitioned into mini-batches and trained over multiple epochs.
We follow the fine-tuning recipe as described in~\cite{green2021} for disordered speech: We start from a speaker-independent  base model trained on typical speech and fine-tune using only the target speaker's data. We only update the first 5 encoder layers (all other weights of the model are frozen), which effectively prevents overfitting.

\subsection{On-device Personalization}
\label{ss:on_dev_pers}

The mobile personalization scenario differs from the server-based scenario in several ways: a) hardware constraints mostly affecting maximum possible memory footprint and training speed, b) limited battery rendering long lasting training processes undesirable, and c) emphasis of reducing the amount of time any private user data will be stored.
\cite{sim19asru} studies on-device personalization for named entities (NEs), showing that by only fine-tuning the RNN-T model's decoder with a relatively small amount of training data, recall of the NEs can be significantly improved. \cite{sim21interspeech} showed that robust continuous on-device personalization can be achieved by having in place a set of acceptance criteria to automatically determine whether a personalized model should be accepted at the end of each training round. To address the data storage effect, \cite{sim19interspeech} studied the effect of having a sliding-window style data consumption pattern for on-device learning. They also compared ways to reduce training memory, including fine-tuning only parts of the model and splitting the gradient computation into smaller blocks.

We here employ a simplification of the sliding-window approach ("consecutive training"), where we collect a small amount of $N$ utterances from a speaker and then train for $E$ epochs with a batch size of $N$. After each such training round, we store the new checkpoint so that we can continue training as soon as the next batch of $N$ utterances has been recorded by the user.
The advantages of this approach are that user data is only kept until a single training round is finished. Moreover, a user can directly benefit from an improved model, which is used to transcribe new phrases as they are recorded. As we show in Section~\ref{ss:consec_training_benefits}, this significantly reduces the user-burden in providing transcribed training examples. 

Even in high-end mobile devices, memory consumption is one of the main limitations for on-device training. The number of layers to be updated has direct implications on memory consumption~\cite{sim19interspeech}. We changed our adaptation recipe to update only encoder layers 2, 3, and 4 leading to a manageable peak memory footprint of 1437 MB (adding more higher layers showed diminishing gains) and a reasonable trade-off between WER degradation and memory consumption\footnote{We chose to update only encoder layers 2, 3 and 4 so that layers 0 and 1 can be computed more efficiently using the quantized weights. Although this results in a 14\% median WER degradation compared to updating layers 0-4, we believe that this is a reasonable compromise.}.

\section{Experimental Results} \label{sec:results}

In this section we present results of our experiments on on-device personalization for disordered speech. We first investigate the best settings for on-device training regarding amount of data needed and base model selection in a server-based training scenario. We then report results of our on-device simulation experiments across 100 speakers with various severities and types of disorded speech.
\subsection{Experimental Settings}

\begin{table}[t]
    \centering
    \caption{Data set of 100 selected speakers, showing median number of test phrases from the home automation (HA) domain, median number of training utterances (across all domains), and sub-sample of training phrases from HA domain.}
    \label{tab:100_speaker_corpus}
    \begin{tabular}{ |c|c|c|c|c| } 
    \hline
    \multirow{2}*{Severity} & No. of & No. of test & \multicolumn{2}{c|}{Train Phrases} \\\cline{4-5}
     & speakers & phrases (HA) & ALL & HA \\
     \hline
     MILD &  48 & 79 & 1,392 & 50 \\ 
     MODERATE & 31 & 80 & 1,648 & 50 \\ 
     SEVERE & 21 & 91 & 1,869 & 50 \\ 
     \hline
    \end{tabular}
\end{table}

We used the Euphonia corpus of disordered speech~\cite{macdonald2021} consisting of over 1 million utterance recordings of over 1000 speakers with different severities and types of speech disorders. All our experiments were performed on a random subset of 100 speakers, who had recorded more than 1000 utterances each. The resulting data set is very diverse, covering 15 different etiologies (31\% with amyotrophic lateral sclerosis (ALS), 21\% Down Syndrome, 13\% Cerebral Palsy, 6\% Parkinson's Disease, etc) and 3 severities of speech impairment (41\% mild, 31\% moderate, 21\% severe).

We used the predefined training, testing, and development splits of the Euphonia corpus. 
For testing, i.e. whenever we are reporting word error rates (WERs), we limit the test sets to utterances from the so-called \textit{home automation} (HA) domain.\footnote{The Euphonia corpus comprises recordings from different domains including home automation queries, caregiver phrases, and longer conversational phases. The home automation subset contains phrases like "turn on kitchen lights" or "play ABBA on Spotify".} For training, we built two sets: a) all training utterances per speaker (spanning all domains) and b) 50 randomly sampled utterances from the speaker's HA training utterances (see Table~\ref{tab:100_speaker_corpus}).

\subsubsection{Base Model}
\label{ss:base_model}

We use a speaker-independent ASR model trained on 162k hours of typical speech \cite{narayanan2018toward} as base model for personalization. This model  has been optimized to (a) be robust across various application domains and acoustic scenarios, and (b) generalize well to unseen conditions. On our data set of 100 speakers with disordered speech, this model yields a median WER of 23.2 on mildly, 41.3 on moderately, and 80.4 on severely disordered speech (see Table~\ref{tab:alldat_vs_50utts}).

\subsubsection{Training}

Server-side training was performed on tensor processing units (TPUs) with the adaptation recipe as described in Section~\ref{ss:on_server_pers}. We use the Adam optimizer with a low learning rate (1e-5) and small batch sizes (32). We train for up to about 100 epochs and perform early stopping based on the hold-out development set to identify the best checkpoint. This adaptation recipe is highly effective, reducing median WER to 3.4 for mildly, 5.2 for moderately, and 9.8 for severely disordered speech when using all data available per user for adaptation (see Table~\ref{tab:alldat_vs_50utts}).

On-device training simulations were run on CPU. We set $N=5$ and $E=4$, and use a learning rate of 1e-3. We run 10 consecutive training rounds as described in Section~\ref{ss:on_dev_pers} to consume a total of 50 utterances from a user. No early stopping is performed here\footnote{We cannot assume a meaningfully large dev set to be available for early stopping  as the on-device scenario is meant to be data-sparse.}, so that WERs reported for on-device training are always based on the \textit{last} checkpoint, i.e. after all 50 utterances have been used. 

\subsection{Tuning the On-Device Training Recipe}
\label{ss:ondev_tuning}

In this section we discuss results of preparatory experiments helping us to determine the best performing and also feasible recipe (given the user and hardware constraints) for on-device based personalization. Experiments reported in this section were conducted in the server-environment.

While ASR personalization with \textit{all available data per user} (amounting to a median of ~1500 utterances) resulted in massive WER improvements, acquiring hundreds to thousand of utterances poses a prohibitively large burden on the user, and this amount of data would also be too much for practical on-device training due to excessive training times on device. In order to understand the potential impact of much smaller amounts of training data, we study the performance of ASR personalization with the small, 50 utterance sub-sample from the HA domain. 

Table~\ref{tab:alldat_vs_50utts} compares WERs of the base model, a model personalized on all of a speaker's training data, and a model personalized on 50 utterances, respectively. While personalization with all training data clearly led to better WERs (median relative WER improvement of 84\%), adaptation with only 50 utterances still shows significant gains (overall median relative WER improvement of 74\%). 

\begin{table}[t]
    \centering
    \caption{Personalization on all vs subsample of 50 utterances compared to base model.}
    \label{tab:alldat_vs_50utts}
    \begin{tabular}{ |c|c|cc|cc| } 
    \hline
    \multirow{2}*{Severity}
    & \multicolumn{5}{|c|}{median WER (rel. impr. over base model)} \\
     \cline{2-6}
     & base model & \multicolumn{2}{c|}{pers all utts} & \multicolumn{2}{c|}{pers 50 utts} \\ 
     \hline
     MILD & 23.2 & 3.4 & (85\%) & 5.8 & (75\%) \\ 
     MODERATE & 41.3 & 5.2 & (87\%) & 11.8 & (71\%) \\ 
     SEVERE & 80.4 & 9.8 & (88\%) & 20.3 & (75\%) \\ 
     \hline
     OVERALL & 33.6 & 5.0 & (84\%) & 8.8 & (74\%) \\
     \hline
    \end{tabular}
\end{table}

The base model was trained on typical speech exclusively (Section~\ref{ss:base_model}). Here, we investigate the impact of using a \textit{seed model} instead -- a model fine-tuned to all speakers of Euphonia's disordered speech corpus. Personalization is then performed starting from this seed model.
For each of the 100 speakers we selected for our experiments, we train a seed model where the \textit{training} data of over 1000 speakers from the Euphonia corpus has been used but the respective target speaker was left out.

Table~\ref{tab:sm_vs_bm_50utts_server} compares server-based personalization performance starting from the base and the seed model, respectively. Personalization from the seed model leads a 9\% increase in WER improvement across all speakers.\footnote{Experiments have shown that the seed model has much lesser impact when larger amounts of training data are available per speaker.}

\begin{table}[t]
    \centering
    \caption{WERs for personalization from base vs seed model with subsample of 50 utts. Last column shows rel WER improvement for personalization from seed model over base model.}
    \label{tab:sm_vs_bm_50utts_server}
    \begin{tabular}{ |c|c|c|c| } 
    \hline
     \multirow{2}*{Severity}
     & \multicolumn{2}{|c|}{median WER } & rel. impr. using \\\cline{2-3}
      & base model & seed model & seed model \\ 
     \hline
     MILD & 5.8 & 5.4 & 7\%  \\ 
     MODERATE  & 11.8 & 10.0 & 15\% \\ 
     SEVERE  & 20.3 & 18.6 & 8\%\\ 
     \hline
     OVERALL  & 8.8 & 8.0 & 9\% \\
     \hline
    \end{tabular}
\end{table}

\subsection{Results of On-Device Training Simulation}

Given our insights from Subsection~\ref{ss:ondev_tuning}, we propose the following personalization procedure for on-device training: a) start from a seed model, b) limit to 50 utterances (more data may be better, but for now we want to study whether 50 utterances will already show significant improvements), and c) update only encoder layers 2, 3, and 4.

\subsubsection{Word Error Rate Improvements}

\begin{figure}[t]
    \centering
    \includegraphics[width=0.5\textwidth]{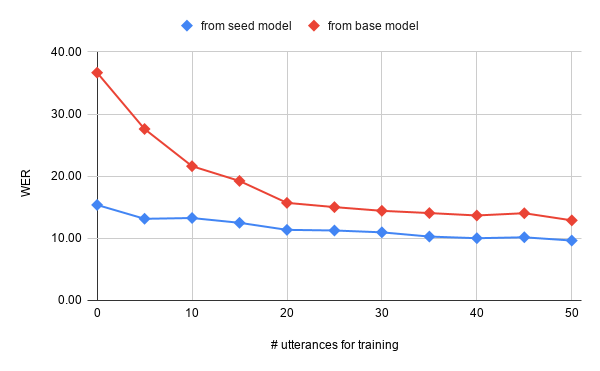}
    \caption{Average WER with increasing number of training for on-device personalization from seed or base model.}
    \label{fig:WER_by_utts}
\end{figure}

Figure~\ref{fig:WER_by_utts} shows how the median WER across all 100 speakers improves as more and more training rounds are performed on device. The biggest improvements happen within the first 4 training rounds ({\em i.e.}, within the first 20 utterances recorded by a speaker). From this graph we can conclude, that further training with additional data beyond 50 utterances would likely lead to some more, but much smaller gains. This figure also highlights once again the importance of the seed model.
Table~\ref{tab:sm_vs_bm_50utts_ondev} shows the final WER after 10 rounds of on-device training. Across all speakers, we can observe a median relative WER reduction of 37\% compared to the seed model, and 71\% compared to the base model. Noticeably, we observe higher improvement rates for speakers with more severe speech disorders ({\em e.g.}, 49\% improvement for severely disordered speech over the seed model, compared to 22\% for mildly disordered speech).

\begin{table}[t]
    \centering
    \caption{WERs and relative improvement of personalized model ("pers") over base model (BM) and seed model (SM).}
    \label{tab:sm_vs_bm_50utts_ondev}
    \begin{tabular}{ |c|cc|c|cc| } 
    \hline
    \multirow{2}*{Severity}  & \multicolumn{3}{c|}{median WER} & \multicolumn{2}{c|}{rel. impr. over} \\\cline{2-6}
    & BM & SM & pers & SM & BM \\ 
     \cline{2-4}

     \hline
     MILD & 23.2 & 8.4 & 6.5 & 22\% & 72\% \\ 
     MODERATE & 41.3 & 18.2 & 12.1 & 34\% & 71\% \\ 
     SEVERE & 80.4 & 37.5 & 19.0 & 49\% & 76\% \\ 
     \hline
     OVERALL & 33.6 & 15.4 & 9.7 & 37\% & 71\% \\
     \hline
    \end{tabular}
\end{table}

\subsubsection{Assistant Task Success Rate}

\begin{table}[t]
    \centering
    \caption{Median Assistant Task Success Rate for base models (BM), seed models (SM), and personalized models (pers).}
    \label{tab:assistant_success_rate}
    \begin{tabular}{ |c|c|c|c|c| } 
    \hline
    \multirow{2}*{Severity}
    & \multicolumn{3}{|c|}{Assistant Task Success Rate}\\\cline{2-4}
     & ~~~BM~~~ & ~~~SM~~~ & pers \\ 
     \hline
     MILD & 52\% & 73\% & 82\%  \\ 
     MODERATE & 38\% & 70\% & 80\% \\ 
     SEVERE & 30\% & 70\% & 82\% \\ 
     \hline
     OVERALL & 40\% & 71\% & 81\% \\ 
     \hline
    \end{tabular}
\end{table}

In all the experiments reported above, we focused on the domain of home automation. Besides WER, a relevant \textit{extrinsic} evaluation metric for this domain is the success rate of correct query intent recognition on the produced transcript. While minor transcription error, like determiners, plurals, probably won't change the outcome, others will completely change the intent, directly deteriorating the user experience.

We here chose the Google Assistant for evaluation. All phrases in our HA test set, if correctly transcribed, are valid Google Assistant queries. We define the \textit{Assistant Task Success Rate} (ATSR) as the percentage of test utterances for which the intent derived from the predicted transcript matches the intent derived from the true transcript. An ATSR of 80\% is generally considered to provide a satisfying user experience.

We randomly chose 20 out of the 100 speakers (5 with mild, 10 with moderate, 5 with severe speech impairment) for which we obtain their ATSR on the base model, the seed model, and the on-device personalized model after all 50 utterances have been used for training. Median ATSRs are shown in Table~\ref{tab:assistant_success_rate}. On-device personalized models attain a median ATSR above the 80\% threshold for all severity groups. It should be noted, that the seed model already has a much improved ATSR compared to the base model, which clearly underperforms with a median ATSR of just 40\% emphasizing how inacceptable speech recognition quality is for users with speech impairment and how this renders voice-controlled services -- like home automation in this case -- 
close to unusable.

Another way to look at the results would be to count the percentage of our 20 speakers hitting the target ATSR of 80\%: 0\% on the base model, 10\% on the seed model, 60\% on the on-device personalized model. Meaning that with just 50 utterances, we are able to provide a useful voice-controlled home automation experience to the majority of the speakers, including those with severe speech impairment.

\subsubsection{User-benefits of consecutive training}
\label{ss:consec_training_benefits}

\begin{table}[t]
    \centering
    \caption{WERs on training utterances comparing \textbf{consec}utive training vs a \textbf{single} training after recording 50 utterances, starting from seed (SM) or base model (SM).}
    \label{tab:consec_training}
    \begin{tabular}{ |c|cc|c| } 
        \hline
        \multirow{2}*{Severity}
        & \multicolumn{3}{|c|}{WER (rel. impr.)}\\
         \cline{2-4}
         & SM single & SM consec & BM consec \\ 
         \hline
         MILD & 10.5 & 8.5  (19\%) & 16.3 \\ 
         MODERATE & 24.6 & 17.0 (31\%) & 30.3 \\ 
         SEVERE & 38.5 & 27.7  (28\%) & 43.0 \\
         \hline
         OVERALL & 20.8& 15.2  (27\%) & 26.2 \\ 
         \hline
    \end{tabular}
\end{table}

Our on-device personalization approach triggers a training round after each batch of $N=5$ utterances has been recorded. Recording is done such that a user speaks a phrase, is then presented with the transcript based on their latest model, and finally corrects any transcription errors before saving the recorded utterance along with the transcript.

A major benefit of consecutive training -- as opposed to a training regime where all data is collected upfront and one single training job is triggered afterwards -- is that users can already benefit from improved models during their recording sessions, leading to reduced transcription errors and hence manual correction work.
Table~\ref{tab:consec_training} quantifies this effect showing the WER over all 50  \textit{training} examples as spoken by the user over all 10 recording iterations as a proxy to transcript corrections needed. We compare on-device personalization with consecutive training against a single training round after all data is recorded. Transcript correction goes down by 27\% across all speakers (median WER of 15.2 compared to 20.8), speakers with moderate speech disorder seeing the biggest gain (31\%).

This analysis also emphasizes another advantage of using seed models -- they also dramatically improve the user experience during recording training utterances by massively reducing transcript correction needs (median WER of 26.2 drops to 15.2 amounting to a relative reduction of 42\%).

\subsection{On-Device Personalization tested in Real Life}

We tested our on-device personalization with one speaker with severely impaired speech (speaker is deaf). Using a Pixel 3 phone, this speaker performed the on-device training exactly as done in our simulations: speaking 50 training phrases across 10 consecutive sessions with the model being updated on-device after each such session. One training round took between 4 and 5 minutes\footnote{This includes all preparatory steps, 
including loading/saving of model checkpoints and model weight quantization for inference.
}, depending on the lengths of the phrases recorded (phrase length was limited to 4 words). Overall, it took the speaker less than 1 hour for both recording and training to be completed (phone was blocked during training to avoid interference with other applications potentially competing for memory and CPU cycles).
Transcript accuracy significantly increased from 48\% (seed model) to 63\% (after personalization).

\section{Conclusions} \label{sec:conclusions}

We have presented an approach for on-device personalization of ASR models with a focus on acoustic adaptation to deal with deviant voice characteristics. Our experiments emphasize the importance of starting from a \textit{seed model}, i.e. a model pre-adapted to a large pool of speakers with speech disorder: it is more effective (leading to bigger WER improvements) and more efficient (requiring fewer transcript corrections) than personalizing from a base model trained on typical speech only.

In simulations on speech recordings from 100 speakers with disordered speech, we showed that with as little as 50 utterances the proposed on-device training procedure is able to reduce the median WER by 71\% and increase the median Assistant Task Success Rate from 40\% to over 80\%. This shows that it is possible to make voice-controlled services, like Google Assistant in this case, usable for people with even severe speech disorder with minimal recording requirement and without the need to transfer any personal data to servers for training.

Future work will include a larger user study to ensure that the promising results we obtained in our simulation will be replicable. Moreover, we plan to investigate on-device personalization for a domain other than home automation, ideally open conversations with longer phrases. This much more challenging adaptation task will likely require more training data and come with increased training times.

\section{Acknowledgements}

We would like to thank Dimitri Kanevsky and Jerry Robinson for testing on-device training with their voice, and Anton Kast and Robert L. MacDonald for early testing and feedback on the on-device training UI.

\bibliographystyle{IEEEtran}

\bibliography{references}

\end{document}